# Broadband High-Performance Terahertz Polarizers by Nanoimprint Lithography for Advanced Applications

ALEXANDRE CHÍCHARO,[1,2,*] ZDENEK KASPAR,[3,4] TATIANA G. RAPPOPORT,[1,5] AJINKYA PUNJAL,[3] CHUN-DA LIAO,[1] PIETER DE BEULE,[1] JÉRÔME BORME,[1] NUNO M. R. PERES,[1,5] AND PEDRO ALPUIM[1]

[1]INL - International Iberian Nanotechnology Laboratory, 4715-330 Braga, Portugal
[2]Humboldt-Universität zu Berlin, Department of Chemistry, 12489 Berlin, Germany
[3]Department of Physics, Freie Universität Berlin, 14195 Berlin, Germany
[4]Faculty of Mathematics and Physics, Charles University, 121 16 Prague, Czech Republic
[5]Center of Physics, University of Minho, 4710-057, Braga, Portugal

*alexandre.chicharo@inl.int

**Abstract:** Terahertz polarizers are needed for advanced spectroscopic systems, but they have drawbacks such as low transmission, short bandwidths, and low extinction ratios. A method for the development of ultrabroadband THz polarizers based on the nanoimprint lithography technique is reported here, in which high performance is achieved for a double-wire-grid polarizer (DWGP) structures on cyclic olefin copolymer (COC) substrates. Over the 0.1 – 25 THz frequency range, the polymer DWGPs had more than twice as much TM-polarized transmittance than their silicon-based counterparts. The degree of polarization was greater than 98% in the 0.1 – 16 THz range, and the extinction ratio was greater than 65.4 dB at 4.2 THz. THz time-domain spectroscopy (THz-TDS) and Fourier-transform infrared spectroscopy (FTIR) were also employed to characterize the optical properties of materials over the frequency ranges of 0.1 – 40 THz and 0.9 – 20 THz, respectively. The nanofabricated polymer DWGP showed better optical properties than the Si DWGP in terms of enhanced TM transmittance and reduced TE leakage. In addition, the prepared COC polarizers exhibited cost-effectiveness, scalability, and durability and can be considered as environmentally friendly alternatives to conventional Si-based polarizers. This study opens up the possibility of using polymer-based DWGPs as important high-performance components in THz imaging and sensing applications and in wireless communication systems.

## 1. Introduction

The terahertz (THz) spectral range, spanning from 0.3 to 30 THz, has long been considered a cutting-edge area in photonics [1–3]. Over the past few years, the field has experienced remarkable progress [4,5]. These advancements have opened up exciting possibilities in areas like spectroscopy, security screening, and wireless communication [5,6]. As a result, the demand for improved THz components—such as emitters, detectors, and passive elements like lenses, filters, and polarizers—is growing rapidly.

Polarizers play a critical role in spectroscopy by allowing only linearly polarized light to pass through while blocking other components [6,7]. When evaluating polarizers, two important factors come into play: the degree of polarization (DOP), which describes how much of an electromagnetic wave is polarized, and the extinction ratio (ER), which indicates how effectively the polarizer separates desired and unwanted polarizations [6].

THz polarizers have been developed using a range of techniques, such as triangular Ag-film gratings [8] , carbon nanotubes [9] , Brewster's angle polarizers [10], liquid-crystal polarizers [11], and wire-grid polarizers (WGPs) [12–18]. WGPs, composed of subwavelength metal wires, can be supported by a substrate or be free-standing [18]. When interacting with THz radiation, conduction electrons in the wires respond primarily along the wire direction, reflecting TE-modes and transmitting TM modes. WGPs are versatile and effective across a wide spectral range, from ultraviolet to THz regions [7]. Their performance is influenced by factors like material selection, geometry, wire pitch, and layer configuration, including single or multiple layers [19–21]. An example is the DWGPs as shown in Fig. 1.

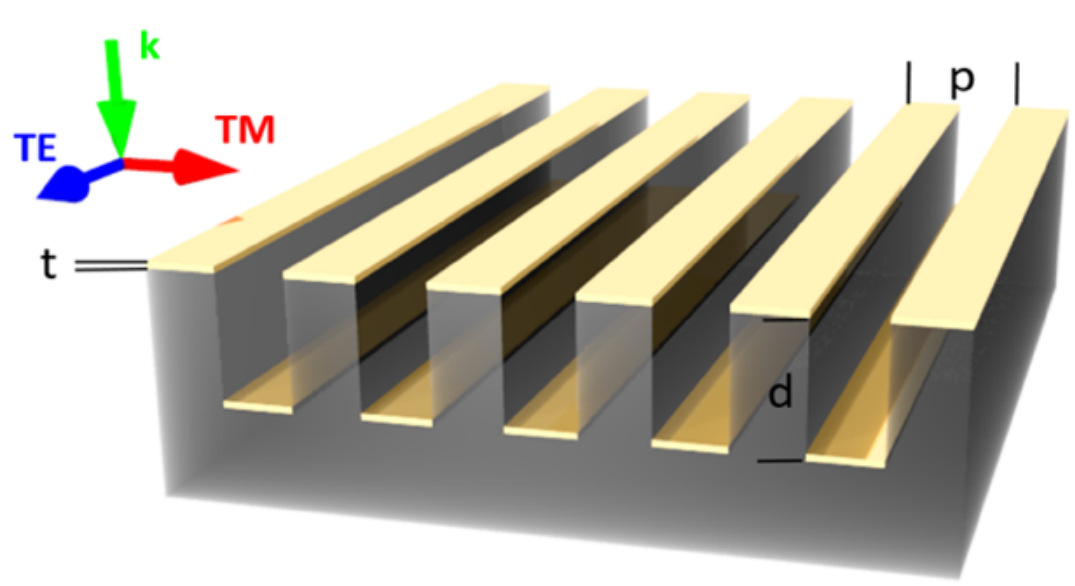

Fig. 1. Schematic of a double-wire-grid polarizer (DWGP) composed of two planes of wires, each forming a single wire grid. Key dimensions include the wire-grid pitch $p$, the spacing between metal layer planes $d$, and the thickness of each metal layer $t$. The fill factor is defined as $p/2$. The directional annotations are defined as the light propagation direction $\mathbf{k}$ (green arrow), direction along the wire grid ($\mathbf{TE}$, blue arrow) and perpendicular to the wire grid ($\mathbf{TM}$, red arrow).

Recent advancements include new materials that enhance ER and novel fabrication methods. Most polarizers use a single plane WGP configuration, while some commercial versions employ free-standing grids, though aligning wires with pitches below 5 µm remains challenging [18]. Techniques like inkjet printing with silver-nanoparticle ink on Kapton film [22] and 3D printing of conductive materials [23] offer rapid production but require further optimization for achieving sub-40 µm pitches. Cleanroom photolithography, combined with sputtering, electroplating, or metal evaporation, enables precise patterning of micron-sized wires [7]. Silicon wafers are commonly used as substrates due to their semiconductor micromachining capabilities and excellent transmission [24]. Other substrates explored include crystalline solids [9,25,26] and polymers [12,27–33].

Single-layer WGP grids achieve an ER of up to 60 dB over a bandwidth of 0.1 - 3 THz [7]. A significantly higher ER can be expected from DWGPs, first proposed in the 1970s. These devices feature aligned grids with sub-wavelength slits, forming a Fabry-Pérot filter with distinct transmission bands [20,21]. Simulation results for this DWGP design show an extraordinary ER of 180 dB [34]. However, fabricating such designs with low pitches and precise alignment is challenging due to increased precision requirements.

A promising DWGP variant, illustrated in Fig. 1, was recently demonstrated using lithography patterning, dry silicon etching, and self-aligned gold deposition. This method achieved a maximum ER of 69.9 dB [19]. Comparative studies indicate that DWGPs exhibit TE transmittance levels much lower than single-layer WGPs, resulting in a threefold increase in ER. This geometry shows significant potential for commercial THz polarizers. However, existing polarizers, particularly those based on silicon, often suffer from limited bandwidth (typically 0.1 – 3 THz), lower TM transmittance, and complex or expensive fabrication processes that hinder scalability.

In contrast, our study presents a novel polymer-based DWGPs using an innovative fabrication approach supported by simulations. These devices were characterized using both THz time-domain spectroscopy (THz-TDS) and Fourier-transform infrared spectroscopy (FTIR) over an ultrabroad THz range. Our results demonstrate the potential of polymers in THz component development, offering enhanced performance and scalability due to their flexibility, lower cost, superior transmittance properties, and more suitability for nanofabrication.

## 2. Materials and Methods

We used two types of polymer films for this study: COC (TOPAS® 6013F04, 140 µm thick, with a glass transition temperature of 138°C) and COP (Zeonor® ZF14-188, 188 µm thick, with a glass transition temperature of 136°C), both sourced from local distributors. Silicon substrates included undoped, intrinsic Si wafers ((100) FZ, double-side polished, 525 µm thick, >10 kΩ·cm, CrysTec GmbH) and p-type boron-doped Si wafers ((100) single-side polished, 725 µm thick, 1–100 Ω·cm, Silicon Valley Microelectronics, Inc.).

The optical properties of the materials and fabricated polarizers were characterized using a THz time-domain spectroscopy (THz-TDS) system, covering a broad frequency range of 0.1–40 THz. The system employed a Ti:Sapphire oscillator operating at an 81 MHz repetition rate, with 15 fs pulse duration and a fluence of 1.2 nJ per pulse, focused onto a spintronic THz emitter. Detection was performed using either a 250-µm-thick GaP ⟨110⟩ or a 10-µm-thick ZnTe ⟨110⟩ crystal, selected for optimal phase-matching at higher frequencies. Measurements were conducted in a dry air environment to eliminate water vapor absorption, and apodization was applied to the time-domain signals to suppress echoes. Assuming $n \gg k$ for a low-absorbing thin film, the real and imaginary parts of the refractive index are calculated as follows [35]:

$$n = \frac{c \, \arg(\hat{T})}{d\omega} + 1 \tag{1}$$

$$k = -\frac{c}{d\omega} \ln\left[\frac{(n+1)^2}{4n} \left|\hat{T}\right|\right] \tag{2}$$

Root-mean-square (RMS) values of the THz time-domain signals were calculated to represent overall performance, and a moving average with a 10-point window was applied to TE data for DOP and ER calculations.

FTIR measurements were conducted using a Bruker Vertex 80v system operating in transmission mode under vacuum. The system included a water-cooled mercury lamp as the far-infrared source, a multilayer beam splitter, and a DLaTGS detector with a polyethylene window. Polarizers were mounted at fixed angles (0°, 30°, 45°, and 60°) on aluminum supports, which were custom-fabricated using CNC milling (as shown in Fig. 4(b)). Data were collected at a resolution of 1 $cm^{-1}$, with each spectrum averaged over 32 scans. Fabry-Pérot fringes were minimized using a moving average filter with a window size of 201 $cm^{-1}$.

Simulations of the polarizers were performed using COMSOL Multiphysics, employing the wave optics module in the frequency domain. TE and TM waves were modeled under normal incidence with Floquet periodic boundary conditions. The refractive indices for Si and COC were set to 3.4 and 1.57, respectively, while minor imaginary components and Fabry-Pérot effects were neglected.

For nanofabrication, silicon molds and substrates were patterned using direct laser writing (Heidelberg DWL 2000) on AZ1505® photoresist. The patterns were transferred to the silicon substrates using inductively coupled plasma-reactive ion etching (ICP-RIE) with an $SF_6$/$C4F_8$ plasma (SPTS Pegasus). Residual photoresist was removed in a plasma asher (PVA TePla GIGAbatch 360M). The thermal nanoimprinting process for the COC substrates was carried out in a vacuum system (Obducat Eitre 8). Si molds treated with an anti-sticking layer (OPTOOL DSX) were heated to 160°C, and a pressure of 30 mbar was applied to imprint the polymer substrates.

Gold deposition was performed using a confocal sputtering system (Kenosistec UHV) with a 15° tilt for uniform coating. A 100-nm-thick gold layer was deposited onto both the nanoimprinted COC and Si substrates. Finally, SEM images were captured with a NovaNanoSEM 650 microscope (5 kV, high vacuum). For cross-sectional imaging, nanoimprinted COC samples were dipped in liquid nitrogen and fractured to create clean cuts for inspection.

## 3. Polarizer materials, design, simulation and fabrication

### *3.1 Materials selection*

### 3.1.1. Substrate material

Silicon is well-known for its optical properties, especially its transparency in the THz region [24,36,37]. Because of its availability, ease of machining, and widespread use in industrial processes, Si substrates are strong contenders for passive THz optical components. However, other inorganic materials such as diamond, thallium bromo-iodide (KRS5), crystal quartz, and sapphire also have good optical properties in the THz range [37]. Despite this, their high cost and reduced THz transmittance—due to factors like reflective losses or birefringence in quartz—make them less ideal for large-scale production.

In recent years, a new class of polymers has shown promising results in the THz region [37–42]. These include polymethyl pentene (TPX) [12], high-density polyethylene (HDPE) [43], polyethylene terephthalate (PET) [27], cyclic olefin polymer (COP), and cyclic olefin copolymer (COC) [40]. PET, while offering a good performance in the THz range, has a higher absorption coefficient compared to COP and COC [39,40], making it less favored for many applications. On the other hand, COP and COC stand out with their superior mechanical properties. These polymers can be manufactured thinner without breaking and are less brittle than Si. Both COP and COC also have low birefringence and are highly resistant to hydrolysis, acids, alkalis, and polar solvents, in addition to offering good gas barrier properties. One commercial version of COP, Tsurupica® (Picarin®), is frequently used in THz lenses [37].

To find the best material for our THz polarizers, we conducted THz transmission measurements for several candidate materials across a broad spectral range. We used THz time-domain spectroscopy (THz-TDS) from 0.1 to 40 THz (Fig. 2) and FTIR transmittance measurements from 0.9 to 20 THz as a secondary method (Fig. S1). The materials tested included two types of Si wafers (undoped/intrinsic and p-doped) and two polymer sheets, COP and COC (see Table 1). The transmission spectra at normal incidence (Fig. 2a) and the extracted THz real and imaginary refractive indices (Fig. 2b and 2c) were measured using THz-TDS.

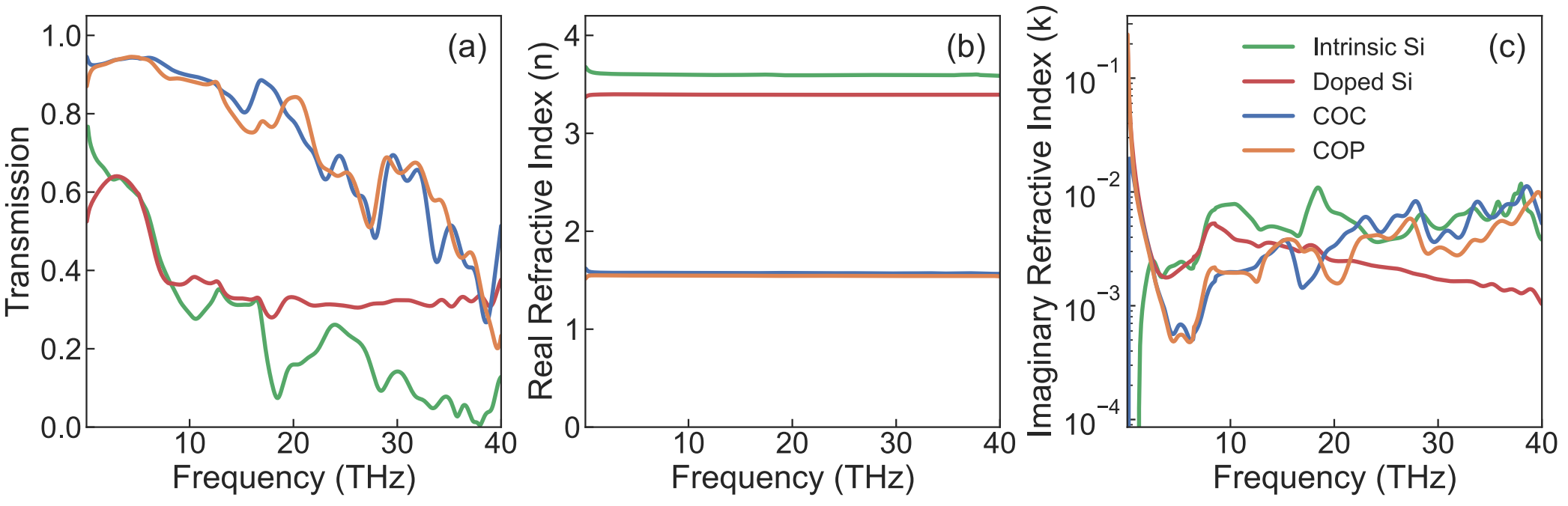

Fig. 2. Characterization of various substrate materials (COC, COP, intrinsic Si, and doped Si) by THz-TDS. (a) Transmission spectra of the materials. (b) Extracted real refractive indices n, and (c) imaginary refractive indices *k* for the different materials.

Both COP and COC show comparable or even better transmission in the THz range than Si across the full spectral range. Intrinsic Si has slightly higher transparency below 4 THz, while p-doped Si performs better than intrinsic Si above 8.5 THz. Additional FTIR measurements (Fig. S1) confirm similar findings in the 0.9 – 20 THz range, showing a consistent absorption peak around 19 THz for Si substrates [36]. The real refractive index (Fig. 2b) of all the materials remains fairly constant across the wavelengths we measured. Both COC and COP have lower refractive indices (~1.5) compared to Si (~3.5), which reduces their reflection losses and indicates less interaction with THz waves—contributing to their higher transmission. These values are consistent with previous research [36,37,44] and further support the suitability of COC and COP for THz applications. Table 1 presents the average real refractive indices of the materials in the 0.1 – 40 THz range.

**Table 1. Materials, material thickness and average refractive index between 0.1 and 40 THz.**

| Substrate Material | Thickness (μm) | Measured Real Refractive Index (*n*) |
|---|---|---|
| COC | 140 | 1.57 |
| COP | 188 | 1.55 |
| Intrinsic Si | 525 | 3.60 |
| Doped Si | 725 | 3.39 |

When looking at the imaginary refractive index (Fig. 2c), intrinsic Si shows very low absorption (<1.5 THz), suggesting minimal absorption at lower frequencies. In contrast, doped Si exhibits much higher absorption at these frequencies, likely due to free carrier effects caused by doping. Both COC and COP have extremely low absorption across the entire 1.5 – 40 THz range, making them ideal candidates for low-loss THz applications. This is particularly important for THz components that require low absorption, stable refractive indices, and flexibility – such as in waveguides, lenses, modulators, and other advanced optical components. In summary, COC and COP films offer much higher transparency across the THz range compared to Si substrates. Because COC demonstrated slightly higher transmission, it was selected for further simulations and polarizer fabrication.

### 3.1.2. Wire-grid material

Choosing the right material for the grid is crucial when building high-performance WGPs. Several options are available, including gold (Au), aluminum (Al), silver (Ag), copper (Cu), indium tin oxide (ITO) [45], carbon nanotubes [25], $VO_2$ [46], and emerging 2D materials [47]. Each of these materials comes with its own strengths. Gold stands out for its excellent conductivity and chemical stability, which make it reliable for consistent performance. Aluminum and silver offer high reflectivity, while copper is a more affordable conductor, though its tendency to oxidize limits its durability. ITO and $VO_2$, as transparent conductive oxides, are valuable for applications requiring transparency, but their thicker films add weight and complicate fabrication. Carbon nanotubes and 2D materials, on the other hand, provide exceptional flexibility and unique electrical properties, although they often lack mechanical strength and face challenges with scalability.

The material choice has a direct impact on the efficiency of TM transmission and TE reflection—both critical for achieving a high extinction ratio (ER) and degree of polarization (DOP) in WGPs. Gold was ultimately chosen because it offers the best combination of conductivity, chemical stability, and ease of deposition. This ensures high TM transmission, long-term durability, and consistent performance across the THz spectrum. As a result, the fabricated polarizers demonstrate excellent polarizing properties while maintaining stability and longevity, making gold-based WGPs a practical choice for demanding THz applications.

### *2.2 Simulation of Polarizers*

To optimize the performance of THz WGPs, we performed simulations of DWGPs using COMSOL (details in Section 2). The two main metrics used to evaluate polarizers are the extinction ratio (ER) and the degree of polarization (DOP) [6]. The ER is calculated as:

$$\text{ER (dB)} = 10\ \log\frac{TM}{\text{TE}} \tag{3}$$

The DOP is given by:

$$\text{DOP}\ (\%) = \frac{TM - TE}{TM + TE} \times 100\% \tag{3}$$

A higher ER indicates a better selection of the desired polarization, leading to improved polarization control. A perfectly polarized wave has a DOP of 100%, while an unpolarized wave has a DOP of 0%.

Our simulations consisted on three key parameters (Fig. 1):

- Grating pitch (p),
- Grating depth (d), and
- Gold layer thickness (t),

with a constant fill factor of p/2. The results (Fig. 3) compare devices made with doped Si (dashed lines) and cyclic olefin copolymer (COC) (solid lines). These simulations highlight the performance of the devices in TM- and TE-polarized modes, as well as the resulting ER. To thoroughly examine the parameter space, we tested multiple configurations. For example, in the first column of Fig. 3, we set d = 2 μm and t = 0.1 μm [Fig. 3(a), (d), (g)]. In the second column, we used p = 2 μm and t = 0.1 μm [Fig. 3(b), (e), (h)], and *p* = 2 μm and *d* = 2 μm in the third column [Fig. 3 (c), (f), (i)].

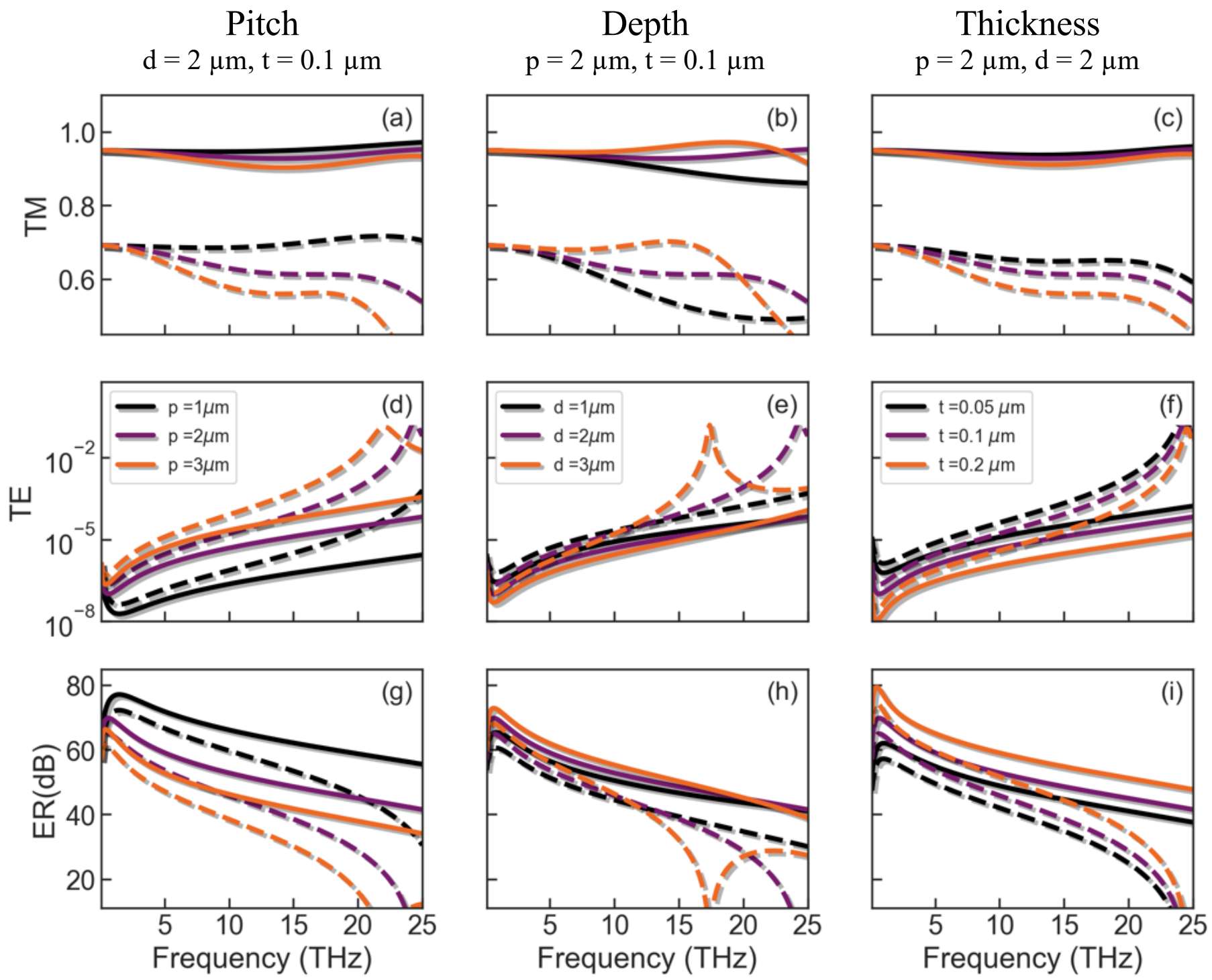

Fig. 3. Simulation results for double-wire-grid polarizers (DWGP) composed of Au wires on COC (solid lines) and doped Si (dashed lines) substrates.

Our simulations clearly show that COC-based devices consistently outperform Si-based devices across a wide range of parameters and frequencies, demonstrating their superiority in the 0–25 THz range. Specifically, COC exhibits > 30% higher TM transmittance compared to Si [Fig. 3(a), (b), (c)], while maintaining consistently lower TE transmittance [Fig. 3(d), (e), (f)]. This leads to a significantly enhanced ER in COC-based devices [Fig. 3(g), (h), (i)]. These findings highlight how critical geometry parameters—pitch (p), depth (d), and gold thickness (t)—impact device performance. By optimizing these parameters, we achieved better TM transmittance, lower TE, and an overall higher ER across the THz frequency range.

Reducing the pitch (p) improved TM transmittance and ER while lowering TE in both materials, leading to more effective polarization. Increasing the depth (d) further enhanced TM transmittance and ER, particularly for COC devices, making them more efficient overall. A thicker gold layer (t) increased ER by reducing TE, though it slightly decreased TM at higher frequencies in Si devices [Fig. 3(i)]. For practical purposes, we maintained a gold thickness of 0.1 μm for a balance between performance and cost.

In summary, optimizing the pitch (p), depth (d), and gold thickness (t) is essential for maximizing the performance of DWGPs, particularly in terms of ER and DOP. Reducing the pitch significantly enhances polarization; however, it introduces challenges in fabrication, reproducibility, and mechanical stability, as narrower trenches and wires are more prone to structural issues. Among the parameters analyzed, depth (d) has the least impact on device performance and should be chosen as a balance between fabrication effort and reliability. Gold thickness plays a more critical role, with thicker layers improving ER by reducing TE transmission. However, this comes at the expense of higher fabrication costs, emphasizing the importance of balancing performance with cost-effectiveness. Our simulations confirmed the superior performance of COC-based DWGPs, which demonstrated higher efficiency and enhanced polarization capabilities compared to Si-based devices.

### *3.3 Nanofabrication of polarizers*

Based on the simulation results, polarizers were fabricated using doped Si and COC substrates through optimized four-step and six-step processes, respectively, as detailed in Fig. 4 (a). These methods ensured precise control over material properties and structural characteristics, tailored for enhanced THz polarization performance. The fabrication begins with Direct-Write-Laser (DWL) patterning of a grating onto a photoresist-coated substrate (Step 1), followed by transferring the pattern through dry etching with $SF_6$/$C4F_8$ plasma using an Inductively Coupled Plasma Reactive Ion Etching (ICP-RIE) tool, achieving groove depths of 2 μm or 3 μm (Step 2). These trench depths were chosen as it optimizes ER while maintaining high TM transmittance, as shown in simulations. Deeper trenches improved TE suppression but introduced fabrication challenges, such as structural instability during nanoimprint soft lithography. After etching, the resist mask is removed with an oxygen plasma asher (Step 3).

For Si-based DWGPs, the Si dies (30 mm x 30 mm) are diced, and a 100 nm Au layer is deposited using a confocal sputtering tool (Step 4a). High anisotropy in the gold deposition is ensured by tilting the sample at a 15° angle, with rotation disabled to achieve optimal alignment towards the target. For COC-based DWGPs, the patterned Si molds are treated with an anti-sticking solution to facilitate demolding (Step 4b). The Si pattern is reusable and transferred to the polymer using soft-lithography in a nanoimprint lithography (NIL) system (Step 5b). The COC sheet is demolded at 50°C, diced, and subjected to the same gold deposition process as the Si substrates (Step 6b).

This combined approach, leveraging both Si and COC substrates with careful process optimization, enables the fabrication of highly anisotropic DWGPs with controlled groove depths and uniform gold coverage. Fig. 4 (b) depicts a COC DWGP on an Al support used in measurements. The selection of materials and optimized fabrication steps ensure high precision, reproducibility, and performance, which are critical for achieving superior THz polarization control. Fig. 4 (c) and 4 (d) show SEM images of the Si and COC polarizers prior to the final Au deposition. Details on sample preparation for the cross-section SEM in Fig. 4(d) can be found in Section 2. Two polarizers of each type were fabricated, with geometries summarized in Table 2.

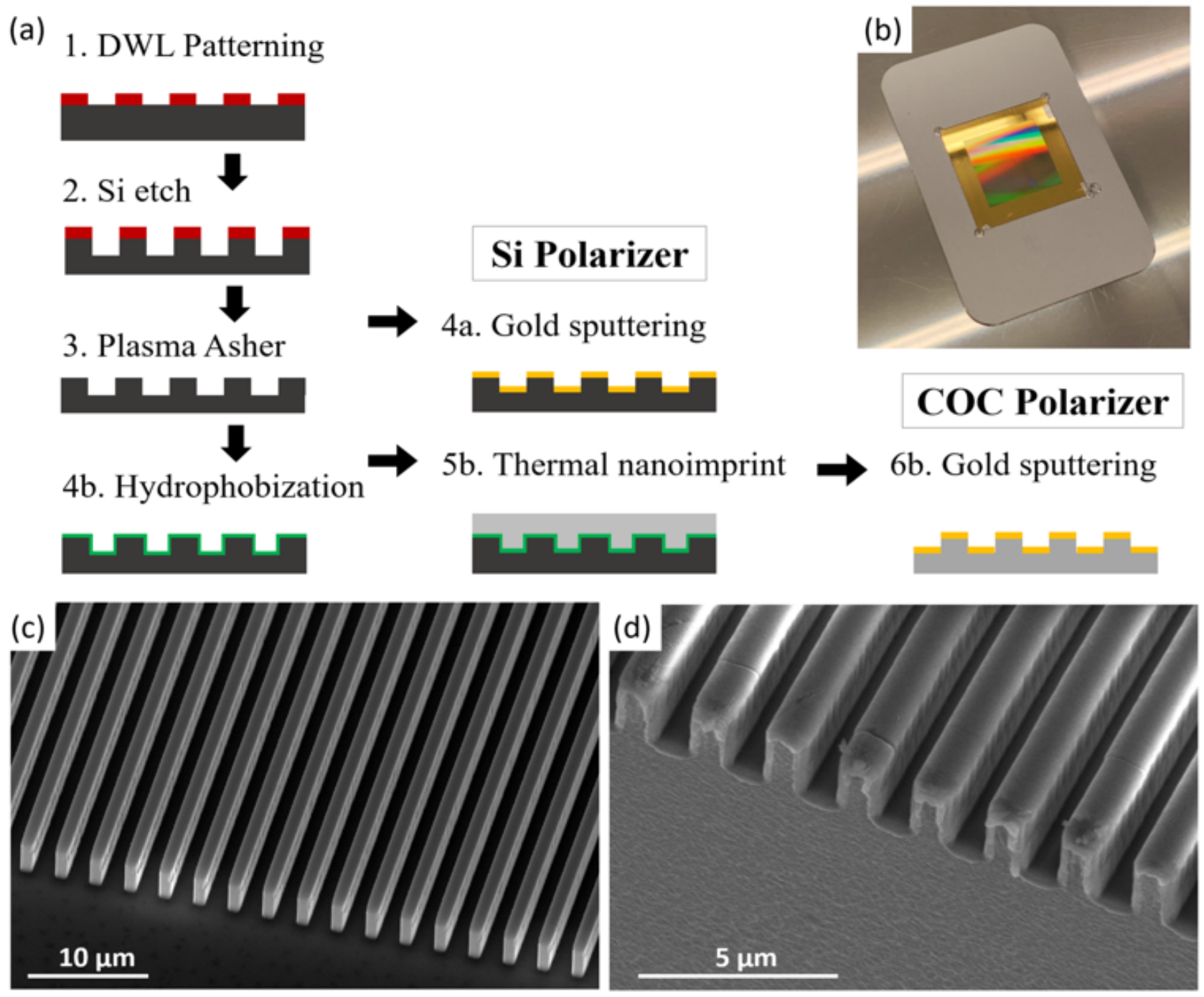

Fig. 4. Nanofabrication schematic of Double-Wire-Grid Polarizers (DWGP). (a) Detailed fabrication steps for Si (black) and COC (grey) double-layer wire grid polarizers, illustrating the sequential use of photoresist (red), a thin hydrophobic agent layer (green), and gold deposition (yellow). (b) COC DWGP after fabrication with Al support for angle measurements. (c) Scanning electron microscope (SEM) image of the Si DWGP after fabrication step 3, emphasizing the uniformity and precision of the wire grid structure. (d) SEM image of the COC DWGP after fabrication step 5b.

**Table 2. Summary of materials and geometric parameters of nanofabricated DWGPs.**

| Polarizer name | Material | Pitch (µm) | Depth (µm) | Gold thickness (µm) |
|---|---|---|---|---|
| P1 | COC | 2 | 3 | 0.1 |
| P2 | COC | 3 | 3 | 0.1 |
| P3 | Doped Si | 2 | 2 | 0.1 |

## 4. Polarizer characterization

### *4.1 Polarizer Performance evaluation via THz-TDS and FTIR*

We evaluated the performance of the fabricated polarizers (P1, P2, and P3) over a broad frequency range using THz-TDS (0.1–25 THz) and FTIR (0.9–20 THz) techniques (see Section 2). These frequency ranges were chosen to capture critical polarizer behavior across the terahertz spectrum, which is relevant for applications like spectroscopy and imaging. While most studies focus on characterizing polarizers in the 0.1–3 THz range, our extended analysis up to 25 THz with THz-TDS and 20 THz with FTIR provides new insights into polarizer performance over an ultrabroadband range. Experiments were conducted with either a single polarizer or a pair of polarizers, as indicated in the relevant figures.

As shown in Fig. 5(a), and consistent with simulation results, the COC-based polarizers outperformed Si-based ones in transmitting the TM-polarized THz electric field, achieving at least a 30% improvement. This superior performance is attributed to the inherent properties of COC, including its lower refractive index and reduced absorption in the THz range. These factors contribute to higher transmittance and overall efficiency. FTIR measurements (Fig. S2(a)) confirmed these observations, showing that Si polarizers consistently exhibit lower TM values than the COC ones. These findings align with the bulk material transmission data presented in Fig. 2 and the simulation results in Fig. 3.

We also analyzed the performance of polarizer pairs, varying the relative angle Θ between them, as shown in Fig. 5(b) and Fig. S2(b, c). In these experiments, the first polarizer was fixed, while the second was rotated by Θ. Malus' law, expressed as:

$$T(\Theta) = T_0 \cos^2(\Theta) \tag{5}$$

where $T_0 = T(\Theta = 0°)$. We accordingly plot a $\cos^2(\Theta)$ function in Fig. 5(b) and found good agreement with the experimental data points obtained after normalization. The normalized experimental data points in Fig. 5(b), closely matched the theoretical $\cos^2(\Theta)$ curve. This strong agreement, further validated by FTIR results (Fig. S2(b, c)), underscores the effective polarization control by the fabricated polarizers.

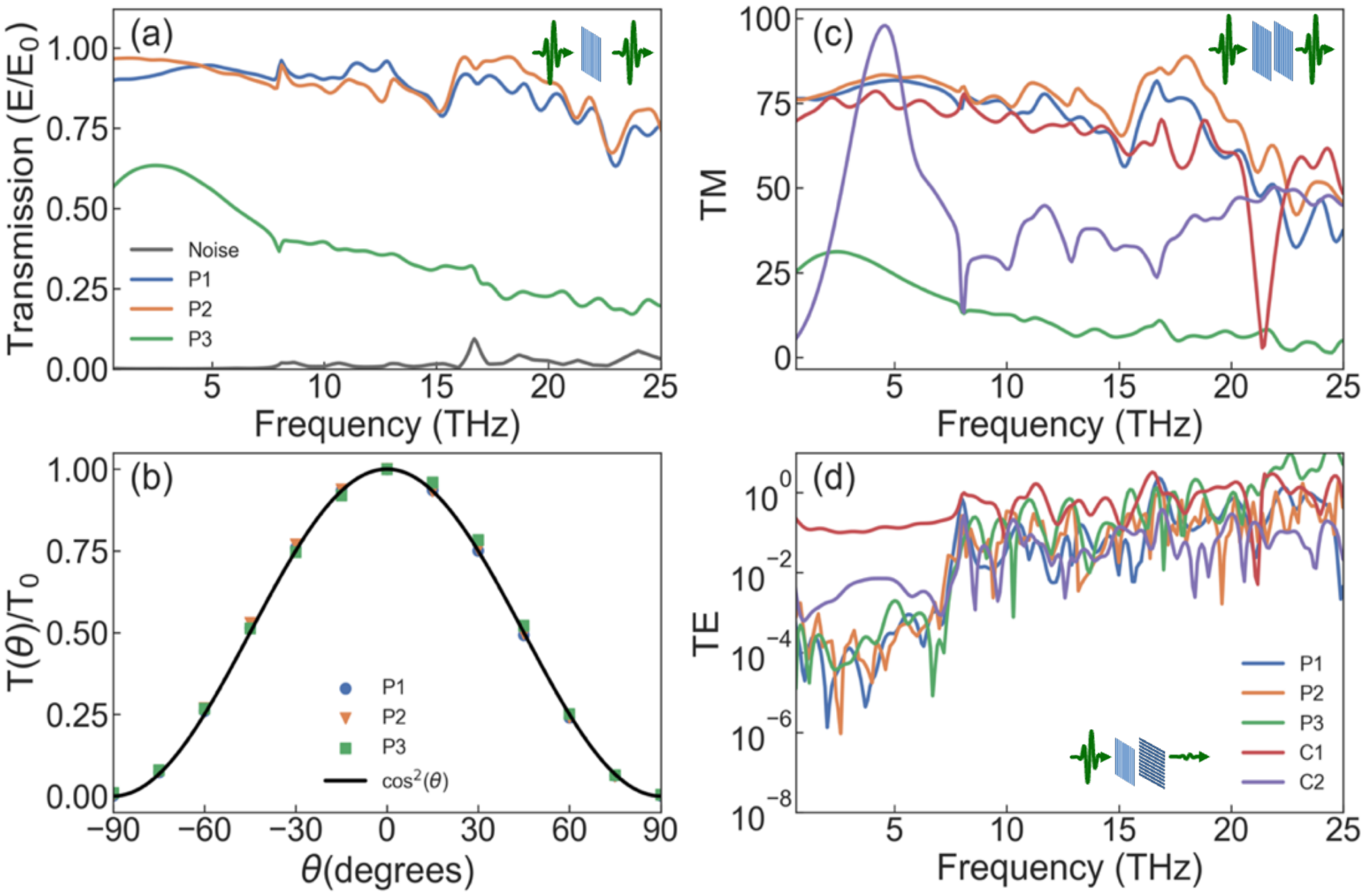

Fig. 5. THz transmission measurements of Polarizers in THz-TDS. (a) Relative THz-field transmission amplitudes of a TM-polarized THz wave through different polarizers: P1) COC substrate with p = 2 µm, d = 3 µm (blue); P2) COC substrate with p = 3 µm, d = 3 µm (orange); P3) Si substrate with p = 1 µm, d = 2 µm (green). Gold thickness for Si- and COC-based polarizes are 0.1 µm. The reference field $E_0$ was recorded without any polarizer. (b) Normalized relative transmitted light intensity vs. relative angle (Θ) between the first polarizer and second polarizer. The solid line shows a $\cos^2(\Theta)$ function. (c) Normalized TM transmittance of each polarizer at parallel alignment (relative angle Θ =0º), and, (d) Normalized TE transmittance at crossed state (Θ =90º).

Figure 5(c) highlights the THz-TDS transmittance ($E^2$) for TM polarization at 0° incidence. To ensure a perfectly polarized THz wave, an additional polarizer of the same type was placed before the 0° polarizer. The results reveal that COC polarizers (P1 and P2) achieved significantly higher TM transmission than Si polarizers (P3), with at least double the transmittance across the THz spectrum. These findings emphasize the advantages of COC substrates in maintaining very high TM transmission.

We also evaluated two commercial polarizers, a free-standing WGP (C1) and a substrate-supported WGP (C2). C1 performed similarly to the COC DWGPs (P1 and P2) but exhibited slightly lower TM values across most frequencies. C2 showed a localized improvement in the 4 – 6 THz range, surpassing all other polarizers, but its TM performance dropped significantly outside this range. Despite these exceptions, both commercial polarizers were generally outperformed by the COC DWGPs in TM transmission. FTIR measurements (Fig. S3(a)) support the THz-TDS results, consistently showing that the COC polarizers exhibited lower attenuation and a more uniform transmittance curve compared to Si polarizers. While simulations predicted higher TM values than observed experimentally – likely due to simplified modeling assumptions such as neglecting full substrate thickness – these trends remained consistent with the key findings of this study.

Figure 5(d) illustrates the TE transmittance ($E^2$) at 90° incidence. TE intensities for both Si and COC DWGPs were comparable but higher than predicted by simulations, likely due to measurement noise affecting TE suppression accuracy. Above 8 THz, all polarizers exhibited an abrupt increase in TE intensity. This is likely an artifact of the system's limited sensitivity to low TE intensities rather than its actual performance. Simulations in Fig. 3, which predicted a more gradual increase in TE due to material losses and geometric constraints, further support this interpretation. Among the commercial options, C1 demonstrated the highest TE levels, especially below 8 THz, indicating weaker TE suppression. In contrast, C2 showed slightly higher TE than Si and COC DWGPs below 17 THz but lower TE at higher frequencies. Overall, the Si and COC DWGPs outperformed both C1 and C2 in suppressing TE transmission.

FTIR measurements in Fig. S3(b) showed similar TE levels for Si and COC DWGPs, which were again higher than simulated values due to the detection noise impacting detection of low TE, similar to THz-TDS. Despite this, the superior TE suppression and TM transmittance of the COC DWGPs confirm their effectiveness for advanced THz applications, outperforming both Si-based and commercial polarizers.

### *4.2 Degree of polarization and polarization extinction ratio*

#### 4.2.1. Degree of polarization

Figure 6(a) shows that the DOP for Si, COC, and C2 polarizers approaches 100% at lower frequencies (<8 THz), highlighting their excellent ability to control polarization in this range. In contrast, the C1 polarizer exhibits lower DOP values at these frequencies, primarily due to higher TE transmission. Notably, the COC polarizers (P1 and P2) maintain exceptional DOP values above 98% up to 16 THz, while Si polarizers (P3) achieve similar DOP levels only up to 9 THz. Beyond this point, the performance of P3 drops significantly, primarily due to a decrease in TM transmittance, making them less effective at higher frequencies. The superior performance of the COC polarizers is largely due to their higher TM transmittance. Since TE transmittance is similar for both COC and Si polarizers, the difference in polarization control is mainly attributed to TM behavior. These characteristics allow COC polarizers to sustain high DOP values over a broader THz frequency range, ensuring better polarization control.

Similarly, the DOP of the commercial C1 polarizer decreases beyond 10 THz. In contrast, the C2 polarizer demonstrates the best overall DOP performance across the entire frequency range, even though it has higher TE and lower TM transmittance compared to the COC and Si DWGPs. This is likely due to the optimized balance between TM and TE intensities in the C2 polarizer, which minimizes polarization leakage and enhances its ability to maintain strong polarization control. Despite these advantages, the COC DWGPs show exceptional consistency, maintaining high DOP values across a wide frequency range and proving their effectiveness for polarization applications.

As simulations indicate, TE measurements were likely influenced by noise, suggesting that the actual DOP values may be higher than reported. To address this, smoothing was applied to the data derived from TE measurements (Fig. 5(d)), as described in the Section 2. This process reduced noise-related fluctuations, improved visualization, and helped to identify overall trends in DOP and ER more clearly.

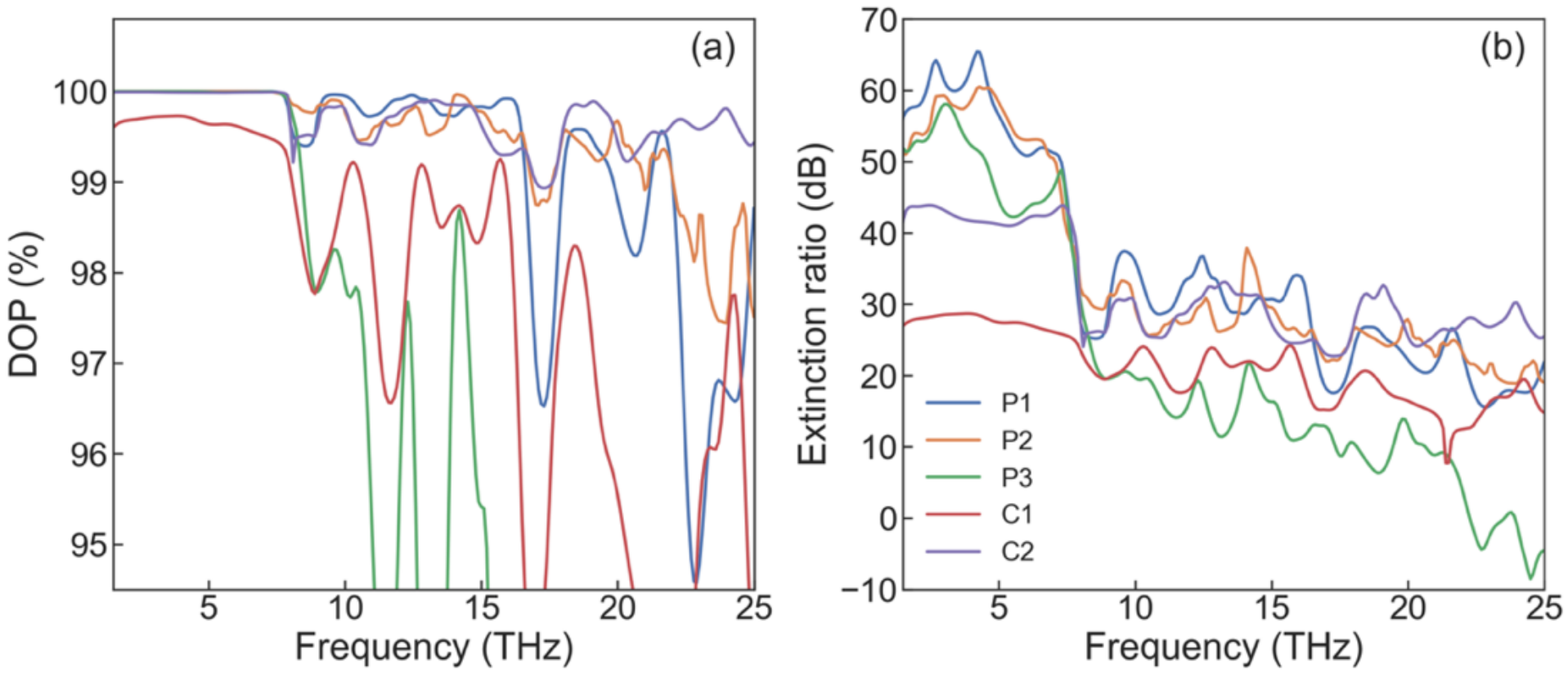

Fig. 6. THz performance measurements of Polarizers in THz-TDS. (a) Degree of Polarization (DOP) and (b) Extinction Ratio (ER) for two COC polarizers (P1 in blue and P2 in orange), one doped Si polarizer (P3 in green), and two commercial polarizers (C1 in red and C2 in purple).

### *4.2.2. Degree of polarization*

Figure 6(b) shows the extinction ratio (ER) for both the fabricated and commercial polarizers, calculated despite challenges caused by noise affecting accurate TE measurements. Beyond 8 THz, the ER values drop abruptly, largely due to increased TE transmission and noise from the detector. In comparison, simulation results in Fig. 3 show a more gradual decline in ER with frequency, driven by material losses and geometric constraints such as pitch (p). This difference suggests that the sharp experimental decline is more a result of system limitations than the devices' actual performance.

The COC polarizers (P1 and P2) consistently exhibited superior ER values across the frequency range, outperforming other polarizers. This advantage is attributed to their stable and higher TM transmittance, combined with lower TE levels. On the other hand, Si polarizers (P3) showed lower ER values due to reduced TM transmission, even though their TE levels were similar to those of P1 and P2. Interestingly, at frequencies above 17 THz, the commercial C2 polarizer slightly surpassed the ER of P1 and P2, likely due to its lower TE and marginally better TM transmission. Among the commercial options, C2 consistently outperformed the free-standing C1 polarizer but still fell short of P1 and P2's performance.

The maximum ER values achieved were as follows: 65.4 dB at 4.2 THz for P1, 60.5 dB at 4.3 THz for P2, 58.1 dB at 3.1 THz for P3, 43.8 dB at 2.4 THz for C2, and 28.7 dB at 3.8 THz for C1. These results highlight the practical advantages of COC polarizers in applications that require precise polarization control. Notably, the actual ER values are likely higher than reported here, as noise in TE measurements likely caused underestimations. Simulations, which predict higher ER values than those measured experimentally, support this interpretation. Despite these limitations, the overall experimental trends align closely with simulations, reinforcing the effectiveness of COC polarizers in achieving high ER values over a broad frequency range.

ER measurements from FTIR, shown in Fig. S3(c), further confirm the superior performance of P1 and P2 compared to P3. While TE levels exceeded predictions from simulations and THz-TDS, P1 and P2 maintained ER values that were at least 4.4 dB higher on average than those of Si polarizers across the frequency range. This improvement is primarily due to the better TM transmittance of COC-based devices. The combined use of THz-TDS and FTIR provided robust validation of the results, offering consistent insights into the optical properties of the polarizers. The agreement between experimental data and simulations further confirms the reliability of the methods used, underscoring the importance of high TM transmittance in enhancing both DOP and ER. These findings demonstrate the effectiveness of the fabricated polarizers for THz polarization control.

Overall, the COC polarizers (P1 and P2) outperformed their Si and commercial polarizers across nearly the entire frequency range. By combining high TM transmittance with reduced TE leakage, they delivered superior polarization control. This strong performance highlights the exceptional potential of COC-based DWGPs for advanced THz applications, highlighting their potential as a valuable components in next-generation technologies.

## 5. Discussion

In this study, we analyzed materials like COC, COP, doped Si, and intrinsic Si over a broad frequency range using THz-TDS and FTIR techniques. By extending the measurements to 0.1–40 THz (THz-TDS) and 0.9–20 THz (FTIR), we addressed a significant gap in the literature, where most studies are limited to the 0.1–3 THz range. This comprehensive analysis highlights the advantages of polymer substrates over Si for advanced THz optical components. The measured refractive indices were consistent with values reported in previous studies, validating the accuracy of our methods and the material properties examined [36,37,41,42,44].

Polymers like COC have shown promise in reducing transmission losses in THz polarizers, as demonstrated in earlier studies [12,14,30,33,48–50]. However, much of the existing work has focused on single-layer WGP configurations with lower ER or on complex multi-layered fabrication processes [33]. Nanoimprint lithography (NIL) has been explored as a method to fabricate single WGPs [12,51–53], with one study achieving a pitch of 140 nm on tri-acetyl cellulose [52]. While this approach offers potential, it is limited by lower ER values. In contrast, our work presents a simpler and more efficient fabrication for polymer-based DWGPs.

Furthermore, our results demonstrate that COC-based DWGPs outperform single-layer configurations and Si-based DWGPs. The design innovations in our approach, including a self-aligning dual wire-grid system, scalable fabrication, and the use of a flexible, transparent, and durable polymer substrate. Notably, COC DWGPs achieved more than double the TM transmission compared to Si-based counterparts. Simulations confirmed that COC DWGPs also exhibit lower TE transmission, leading to higher DOP and ER values, making COC DWGPs a strong candidate for advanced THz polarization control.

Our simulations further suggest that performance could be enhanced by refining design parameters, such as reducing pitch, increasing trench depth, and thickening the gold layer. For example, a single Si DWGP achieved an ER of ~69.9 dB at 1.67 THz by using a thicker gold layer (4× thicker than in our work) and a larger pitch ($p = 4$ μm, $d = 5$ μm, $t = 0.4$ μm) [19]. Another study achieved an average ER of 40 dB within 0.5–2 THz using thinner gold and shallower trenches ($p = 4$ μm, $d = 1$ μm, $t = 0.1$ μm) [17]. These findings align with our results, emphasizing that smaller pitches, deeper trenches, and thicker gold layers improve ER. Dual-layer DWGP designs have also shown potential, achieving ER values as high as 87 dB at 1.06 THz by patterning both sides of a Si substrate [17]. Applying a similar dual-sided approach to COC DWGPs using NIL and gold sputtering could further enhance performance while maintaining the scalability and flexibility of polymers.

This study uniquely combines THz-TDS and FTIR to provide a comprehensive broadband characterization of polarizers, covering 0.9–20 THz. Additionally, we compared our fabricated polarizers directly with commercial options—an analysis rarely addressed in the literature. Our findings demonstrate that COC DWGPs offer cost-effectiveness and scalability compared to Si-based alternatives. The NIL fabrication process is not only faster but also more environmentally friendly and scalable than traditional methods like direct-write lithography or HF etching used for Si materials. Moreover, the COC polarizers exhibit a broader operational frequency range, making them suitable for various applications.

The flexible nature of the polymer substrate offers significant advantages over brittle Si, including compatibility with roll-to-roll manufacturing and curved surfaces [54]. Combined with their exceptional TM transmittance and reduced TE leakage, COC DWGPs deliver high-performance polarization control for advanced THz applications. Our findings validate the strong performance and scalability of COC DWGPs, setting them apart from traditional Si-based designs. These unique properties pave the way for cost-effective THz components, such as filters, modulators, and waveguides, and position COC DWGPs as a transformative solution for next-generation THz systems.

## 6. Conclusion

Our study pushes the boundaries of terahertz (THz) polarizer technology with the development of Double-Wire Grid Polarizers (DWGPs) on flexible cyclic olefin copolymer (COC) substrates. Using nanoimprint soft-lithography, we achieved subwavelength pitches of 2–3 μm and trench depths of 3 μm, showcasing significant advancements in nanofabrication processes for THz applications.

Compared to traditional silicon (Si) polarizers, COC DWGPs demonstrated more than double the transmittance of TM-polarized light across the 0.1–25 THz spectrum. They also exhibited higher degrees of polarization (DOP) and extinction ratios (ER). Through comprehensive characterization using THz-TDS (0.1–40 THz) and FTIR (0.9–20 THz), we highlighted the superior optical properties of COC over COP, intrinsic Si, and doped Si substrates. The fabricated COC DWGPs achieved outstanding TM transmittance, reduced losses, and consistent polarization performance, aligning closely with finite-difference time-domain (FDTD) simulations and confirming the validity of our approach.

This work highlights the importance of materials science in advancing THz technologies. By optimizing DWGP nanostructures and leveraging the unique advantages of COC substrates, we demonstrated clear benefits over traditional Si-based designs. Beyond their excellent optical performance, COC DWGPs offer cost-efficiency, durability, and environmental sustainability. Their compatibility with roll-to-roll manufacturing and curved surfaces further underlines their potential as a potential solution for next-generation THz optical components.

**Funding.** This work was supported by the Portuguese Foundation for Science and Technology (FCT) through projects GRAPHSENS (POCI-01-0145-FEDER-028114; PTDC/FISMAC/28114/2017) and TARGET (UTA-EXPL/NPN/0038/2019). T.G.R. acknowledges funding from Instituto de Telecomunicações—grant number UID/50008/2020 in the framework of the project Sym-Break. Z.K. acknowledges support from the Horizon 2020 Framework Program of the European Commission under FET Open Grant No. 863155 (s-Nebula) and the support by TERAFIT project No. CZ.02.01.01/00/22_008/0004594 funded by OP JAK, call Excellent Research.

**Acknowledgment.** We acknowledge the Terahertz Physics Group at Freie Universität Berlin, in the labs of which terahertz measurements were performed.

**Author Contributions.** A.C.: conceptualization, methodology, software, validation, investigation, resources, data curation, writing the original draft, reviewing and editing manuscripts, creating visualizations, supervision, project administration, and funding acquisition. T.G.R., Z.K., A.P.: methodology, software, validation, investigation, resources, data curation, writing the original draft, reviewing and editing manuscripts, creating visualizations. C.D.L., P.D.B.: writing - reviewing and editing manuscripts. J.B., T.K., N.M.R.P., P.A.: validation, investigation, resources, reviewing and editing manuscripts, supervision, project administration, and funding acquisition.

**Disclosures.** The authors declare no conflicts of interest.

**Data availability.** Data underlying the results presented in this paper are not publicly available at this time but may be obtained from the authors upon reasonable request.

# Supplemental Information: Broadband High-Performance Terahertz Polarizers by Nanoimprint Lithography for Advanced Applications

## S1. FTIR Transmission spectra of various substrate materials

We characterized selected materials using Fourier-transform infrared spectroscopy (FTIR) equipment with a Far-IR input source in transmission mode under vacuum conditions (see Section VII C). Figure S1(a) shows the high-resolution spectra (1 $cm^{-1}$) of all these materials, which exhibit a characteristic fringe pattern consistent with classical electromagnetic theory [36]. To illustrate the overall transmittance, an average was taken across the spectra. Figure S1(b) provides a detailed view of the fringe pattern near 10 THz. These fringes result from Fabry-Pérot interference, caused by multiple reflections between the two reflective surfaces of the material, arising from constructive and destructive interference between transmitted light and light undergoing multiple internal reflections.

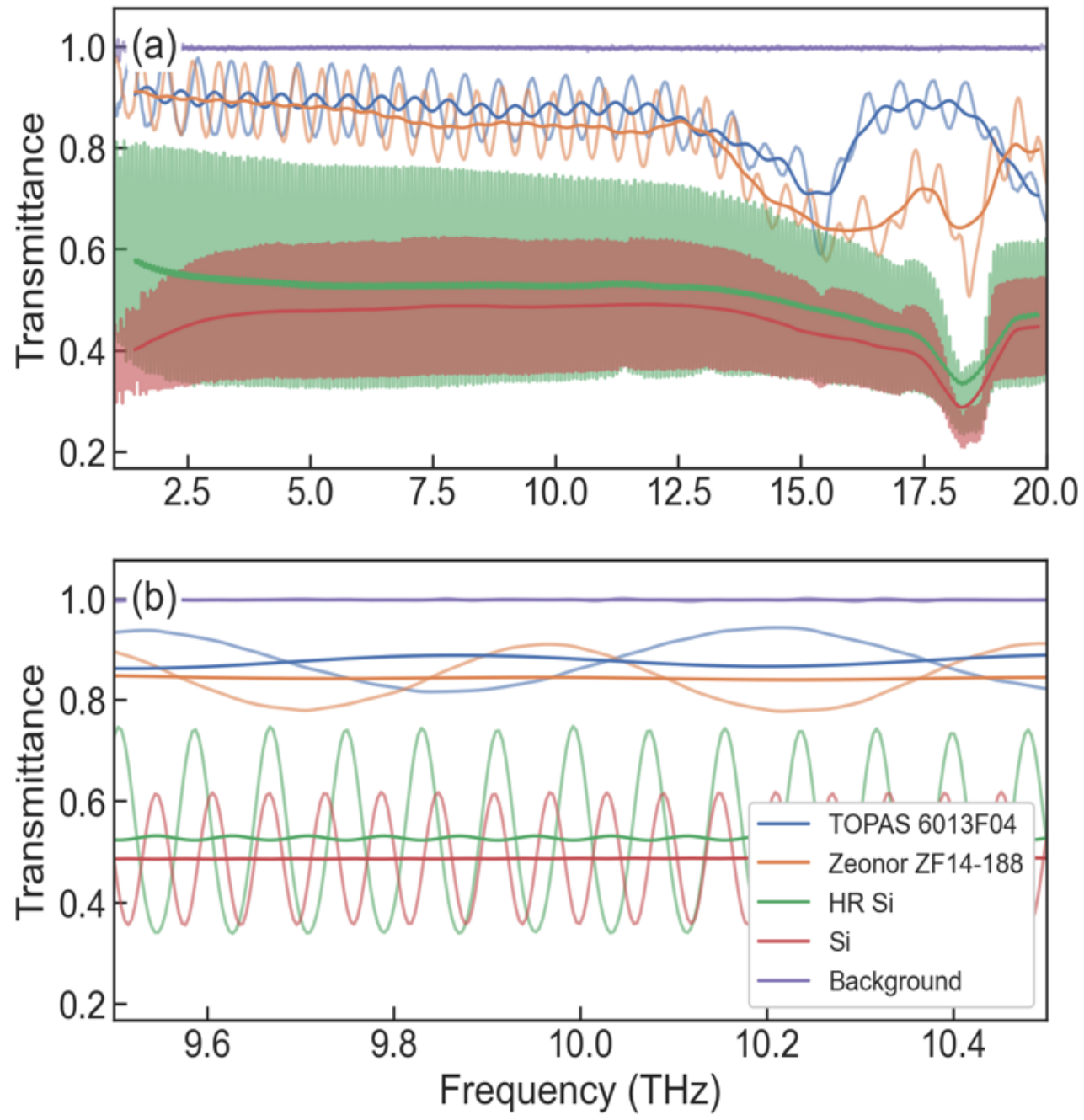

**Fig. S1(a)** FTIR transmission spectra of intrinsic Si, doped Si, COP, and COC polymers. The spectra are shown with transparent fringe lines, while the solid lines indicate the moving geometric mean of the measured fringe patterns for each material, color-coded accordingly. **(b)** A detailed zoom-in of the average transmission and fringe pattern of each material near the 10 THz region, highlighting subtle variations in performance and material-specific behavior.

The behavior of light upon entering the substrate is complex: some rays pass directly through, while others undergo internal reflections, causing a phase shift and reversal of direction at each substrate/air interface. When destructive interference occurs, the phases of all exiting rays cancel each other out, resulting in a band of lower intensity, whereas constructive interference leads to higher intensity. Importantly, the observed fringe pattern does not significantly impact the performance of the polarizers. Spectroscopic techniques account for and remove such background patterns from system components, including variations in light intensity over time, internal reflections, and other artifacts. This ensures that the final performance metrics accurately reflect the polarizer's capabilities without interference artifacts.

Moreover, it is well known that the thickness of the films ($t_F$) can be obtained from analyzing the fringe pattern. The thickness is inversely proportional to the mean distance between interference maxima (Δf) within a selected frequency range and depends on the refractive index (n) of the film, which is assumed constant in that interval: $t_F = 1/(2.n.\Delta f)$. [55] Thicker materials and higher refractive indices result in denser fringes, as observed in the spectra of both Si wafers. The substrate thicknesses were calculated using FTIR software OPUS v8, assuming refractive indices of 3.4 for Si and 1.5 for COC/COP substrates [36,37,46] over the frequency range of 3 to 12 THz. The calculated thicknesses ($t_F$) are: 729.9 µm and 541.3 µm for doped and HR Si, respectively; 190.7 µm for COP, and 154.6 µm for the COC substrate. These values are close to the reference thickness values presented in section 5.A. Additionally, differences in substrate polishing can explain the lower fringe amplitude in the doped Si sample compared to the HR Si sample: doped Si is single-side polished (SSP), which reduces internal reflection compared to undoped/HR double-side polished (DSP). COP and COC substrates, being thinner and having lower refractive indices, exhibit lower frequency fringes compared to Si samples.

## S2. FTIR characterization on polarizers

We characterized our fabricated polarizers using FTIR equipment with a Far-IR input source in transmission mode under vacuum conditions (Section VII C.). Four different polarizers were evaluated: P1 (COC, p = 2 µm, d = 3 µm), P2 (COC, p = 3 µm, d = 3 µm), and P3 (Si, p = 1 µm, d = 2 µm). As shown in Figure S2(a), the transmitted light from both COC polarizers outperformed both the Si polarizers. The COC polarizers exhibited at least twice the transmittance polarized light compared to Si polarizers, which is particularly significant for many THz systems and applications due to the inherently low radiation amplitude. Notably, the lower pitch COC polarizer (p = 2 µm) displayed slightly higher transmittance than the larger pitch version (p = 3 µm), aligning well with the simulation results. In comparison, Si polarizers showed less than half the transmittance efficiency of COC polarizers, reflecting similar behavior to the bulk material transmittance characteristics seen in Figure S2(a).

Since the light from the FTIR source was unpolarized, additional experiments were conducted using two polarizers set at various relative angles (0°, 30°, 45°, 60°, 90°), using fabricated Al supports, to evaluate the performance of the fabricated devices according to Malus' law. For this purpose, one COC polarizer (P1: p = 2 µm, d = 3 µm, t = 0.1 µm) was mounted in the FTIR equipment, and a background measurement of the light transmitted by this polarizer was recorded ($T_0$). Figure S2(b) shows the average light transmission of a second COC polarizer (P1: p = 2 µm, d = 3 µm, t = 0.1 µm) at different relative angles to the first polarizer. At Θ = 0°, maximum transmission was observed, with a slight decrease due to material damping. This represents the transmitted TM component of the polarizer. At Θ = 90°, the transmitted light was nearly zero, as expected for two crossed polarizers, representing the transmitted TE component. As seen in Fig. S2(b), the transmission decreased for all frequencies as the angle Θ increased, reaching a minimum at 90°, thereby confirming broadband polarizer functionality from 0.9 - 20 THz. A fitting to Malus' law was performed at 10 THz for all polarizers, with results shown in Figure S2(c). These results confirm that all polarizers adhere to Malus' law, indicating the high quality and effectiveness of our fabricated DWGPs.

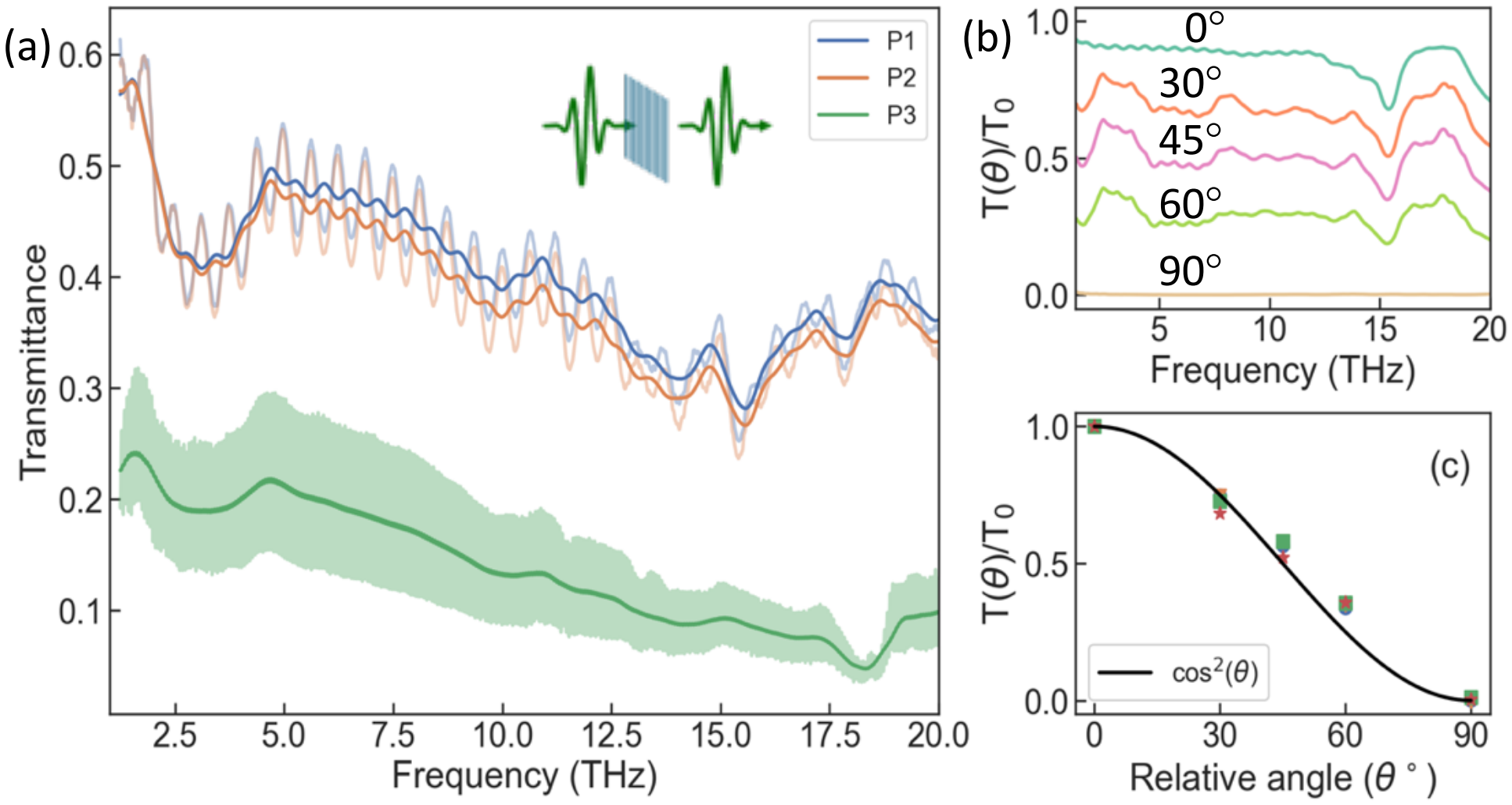

**Fig. S2** THz transmittance measurements of Polarizers in FTIR. **(a)** Relative transmittance intensity of single polarizers: P1 [COC with pitch (p) = 2 μm and depth (d) = 3 μm (blue)], P2 [COC with p = 3 μm and d = 3 μm (orange)], and P3 [Si with p = 1 μm and d = 2 μm (green)]. The gold thickness for both Si and COC polarizers is 0.1 μm. **(b)** Normalized transmittance of COC with p = 2 μm and d = 3 μm as a function of relative polarizer angles (Θ). **(c)** The corresponding fitting curves to $\cos^2(\Theta)$ at 10 THz. Color coding is consistent with Fig. 5(a).

## S3. FTIR extinction ratio of polarizers

To evaluate the extinction ratio (ER) of our Si and COC DWGPs, we measured TM and TE transmissions using two polarizers positioned at relative angles of Θ = 0° and Θ = 90°, respectively, as shown in Fig. S2. Polarizer P1 (COC, p = 2 μm, d = 3 μm, t = 0.1 μm) was utilized to generate the incident polarized beam. Figure S3 illustrates the TM light transmission of the four polarizers, revealing an overall decrease in intensity due to substrate absorption and reflections. Notably, COC polarizers demonstrated superior transparency across the entire broadband range, with the smaller pitch (p = 2 μm) outperforming the larger pitch (p = 3 μm). While Si polarizers covered the full range, they exhibited significantly lower transparency compared to COC polarizers, which consistently delivered at least double the transmission intensity of Si.

TE measurements at Θ = 90° were processed using a moving window geometric mean to mitigate noise fluctuations. As shown in Fig. S3(b), the TE intensities for all polarizers were similar but higher than predicted by simulations, indicating that these TE signals were predominantly noise from the FTIR detector. Simulations in Fig. 3 predicted transmission intensities below $10^{-3}$, which is much lower than the values measured in Fig. S3(b) or observed in similar studies involving single WGPs on COC films.16 The limitations of the FTIR system became apparent, as similar results were observed when incident light was fully blocked, suggesting that the system lacks the sensitivity to accurately measure very low TE transmissions. This limitation impacts the precision of the ER values obtained for these devices with FTIR.

Consequently, we infer that the actual ER is likely higher than the values shown in Fig. S3(c). COC polarizers exhibited an average ER improvement of at least 4.4 dB across the full range compared to Si polarizers, which is attributed to their superior transparency. Since ER is directly proportional to the TM/TE ratio, the enhanced TM transmission of COC polarizers resulted in better performance and higher ER values compared to Si polarizers across the entire frequency range. This trend aligns well with the predicted ER values shown in Fig. 3(g, h, i), reinforcing the superior capabilities of COC polarizers for THz applications.

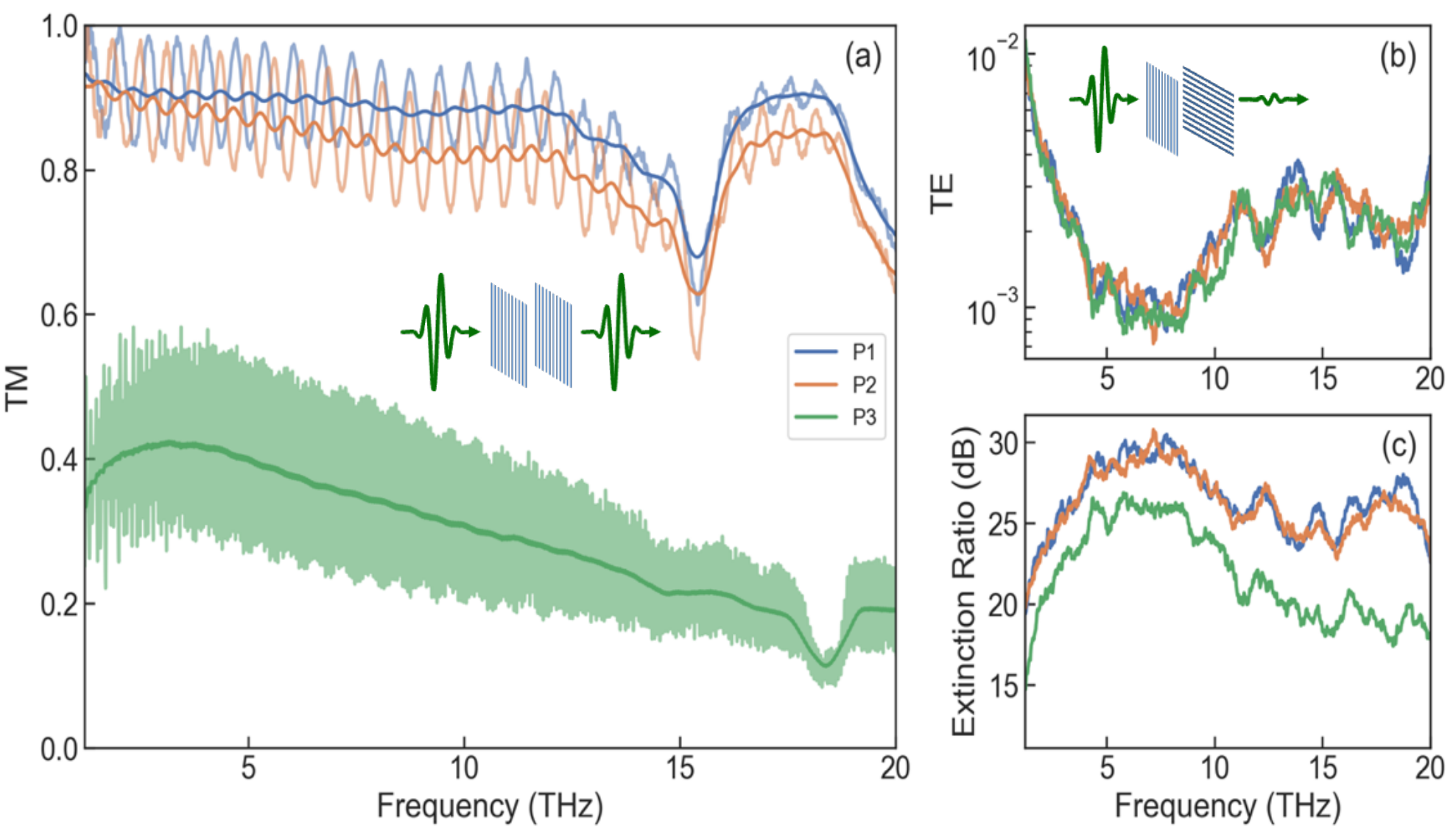

**Fig. S3** THz performance measurements of Polarizers in FTIR. **(a)** Relative transmittance TM intensity of polarized light by polarizers: P1 [COC with pitch (p) = 2 µm and depth (d) = 3 µm (blue)], P2 [COC with p = 3 µm and d = 3 µm (orange)], and P3 [Si with p = 1 µm and d = 2 µm (green)]. The gold thickness for both Si and COC polarizers is 0.1 µm. **(b)** Relative averaged TE transmittance in crossed configuration for the same polarizers. **(c)** Calculated extinction ratio (ER) for polarizers, demonstrating the varying degrees of polarization efficiency across the different polarizer designs.